\documentstyle[]{l-aa}

\def\cg/{\mbox{2CG 135+01}}
\def\gts/{\mbox{GT 0236+610}}
\def\groname/{\mbox{2EG J0241+6119}}
\def\qso/{\mbox{4U 0241+61}}
\def\lsi/{\mbox{LS I $+61^\circ303$}}
\def\ergss{\hbox{ ergs s}^{-1}}

\def\photcmsmev{\hbox{ photons cm}^{-2}\hbox{ s}^{-1}\hbox{ MeV}^{-1}}
\def\ergss{\hbox{ ergs s}^{-1}}

\def\solmas{\hbox{ M}_\odot}
\def\decree#1#2{#1#2^{\circ}}
\def\phibar{\bar\varphi}
\def\phigeo{\varphi_{g}}
\def\molh{\hbox{H}_2}
\def\etal/{et al.}
\def\asterisk{\lower 3pt\hbox{*}}
\def\gamms/{$\gamma$}

\begin{document}

\thesaurus{13.07.2 ; 08.09.2 \cg/ ; 08.09.2 \gts/}
\title{COMPTEL detection of the high-energy \gamms/-ray source \cg/}

  \medskip
  \author{R. van Dijk\inst{2,5,}\thanks{Present
     address: Astrophysics Division, ESTEC, Noordwijk.}
	\and	K. Bennett	\inst{4}
	\and	H. Bloemen	\inst{2}
	\and	W. Collmar	\inst{1}
        \and    A. Connors      \inst{3}
	\and	R. Diehl  	\inst{1}
	\and	W. Hermsen	\inst{2}
	\and	G.G. Lichti	\inst{1}
        \and    M. McConnell    \inst{3}
        \and    R. Much         \inst{4}
	\and	V. Sch\"onfelder\inst{1}
        \and    H. Steinle      \inst{1}
        \and    A. Strong       \inst{1}
	\and    M. Tavani	\inst{6}
  }

  \institute{
               Max-Planck Institut f\"ur Extraterrestrische Physik,
               P.O.~Box 1603, 85740 Garching, F.R.G.
  \and
               SRON-Utrecht,
               Sorbonnelaan 2, NL-3584 CA Utrecht, the Netherlands
  \and
               Space Science Center, Univ.~of New Hampshire,
               Durham NH 03824, U.S.A.
  \and
               Astrophysics Division, ESTEC, P.O.~Box 299,
               NL-2200 AG Noordwijk, the Netherlands
  \and
               Astronomical Institute, Univ.~of Amsterdam,
               Kruislaan 403, NL-1098 SJ Amsterdam, the Netherlands
  \and
               Columbia Astrophysics Laboratory, New York NJ 10027, U.S.A.
  }

  \date{Final version sent to A\&A on April 19, 1996}

  \offprints {R. van Dijk (rvdijk@astro.estec.\allowbreak esa.nl).}

  \maketitle

  \begin{abstract}
We present an analysis of COMPTEL observations made between November 1991
and May 1994 of \cg/, a bright \gamms/-ray source located near the
Galactic plane.
At energies above 1 MeV, an excess consistent
with the position of \cg/ is detected in the sum of the observations, at flux
levels which are a factor of $10-100$ below those published in the past.
The detection significance of this excess, when the possible presence
of underlying Galactic diffuse emission is neglected, is $6.6\sigma$
for 3 degrees of freedom.
The differential photon spectrum in the 1--30 MeV energy range can
be described by a power law with a spectral index of $1.95^{+0.2}_{-0.3}$. Due
to the uncertainties involved in modelling the Galactic-disk diffuse emission
underneath the source, the absolute flux levels must be considered uncertain
by a factor of two. They are consistent with the
extrapolation of the time-averaged spectrum of \cg/ measured with EGRET,
thereby strengthening the identification. No significant temporal correlation
between the \gamms/-ray emission and the monitored
radio emission of the possible counterpart radio source \gts/ (showing a
26.5 day modulation) is found.


  \end{abstract}

\section{Introduction}

With a flux above 100 MeV of $1.0\times10^{-6}$ photons cm${}^{-2}$ s${}^{-1}$,
the \gamms/-ray
source \cg/ is one of the brightest unidentified high-energy
Galactic-plane sources in the second COS-B catalogue (Hermsen \etal/ 1977;
Swanenburg \etal/ 1981).
\cg/ attracted attention when its position was found
to be consistent with that of the radio source \gts/,
which exhibits strong radio outbursts of a non-thermal character
(Gregory \& Taylor 1978). Long-term radio observations of \gts/ subsequently
revealed a 26.496-day periodicity in these
outbursts (Taylor \& Gregory 1982, 1984), of which the amplitude is possibly
modulated by a 4-year period (Gregory \etal/ 1989; Paredes \etal/ 1990).
The decay of the radio flux during outbursts is reminiscent of synchrotron
emission from an expanding cloud of relativistic electrons (Taylor \& Gregory
1984). In a high-resolution map obtained two days post-outburst from VLBI
measurements at 6 cm, the object appears to consist of two components
separated by
$3.1\times10^{13}(D/2.3\hbox{ kpc})\hbox{ cm}$ (Massi \etal/ 1993).
Optically, \gts/ has been identified with \lsi/, which
has a spectrum typical of an early-type B star (Hutchings \& Crampton 1981;
Maraschi \etal/ 1981; Paredes \& Figueras 1986) and exhibits the near-infrared
excess commonly observed for Be-type stars (D'Amico \etal/ 1987;
Hunt \etal/ 1994).  Recent modelling of the JHK-band light curves of \lsi/
has shown that the onsets of the radio outbursts roughly coincide with
the inferred periastron passage (Mart\'\i\ \& Paredes 1995).
The source has also been detected in X-rays (Bignami \etal/ 1981;
Goldoni \& Mereghetti 1995). The X-ray spectrum is rather hard compared to
those of normal B stars, but is similar to spectra observed for Be/X-ray
binaries.
\medskip

In recent observations made with EGRET on board the Compton Gamma-Ray
Observatory (CGRO),
a source designated \groname/ was detected
on several occasions (Thompson \etal/ 1995). This source, whose position is
consistent with that of \cg/ and which is roughly as bright, is generally
assumed to be the COS-B source.
The improved position of the \gamms/-ray source established by the EGRET
detection of \groname/ (a 95\% error radius of $33'$) is consistent
with that of \gts/ (located at $(l,b)=(\decree{135}{.68},\decree{1}{.09})$)
and has
strengthened the proposed identification with the radio source. In addition,
the nearby QSO \qso/ (located at $(l,b)=(\decree{135}{.64},\decree{2}{.43})$),
which was also contained in the relatively large COS-B error region, is
now firmly rejected (von Montigny \etal/ 1993).

A detection of the 26.496 day periodicity in \gamms/-rays would
remove any remaining doubt about the identification with \gts/, but
neither COS-B nor EGRET has yet detected such flux variations.
Note that the 26.496-day periodicity inferred from radio observations
of \gts/ has possibly been detected at infrared and optical wavelengths as well
(Mendelson \& Mazeh 1989; Paredes \etal/ 1994),
with smaller flux variations towards shorter wavelengths.
At UV wavelengths and in X-rays, where the positional accuracy is good
enough to allow a confident identification with \gts/, orbital-phase related
flux variations have not been detected up to recently
(Bignami \etal/ 1981; Howarth 1983; Goldoni \& Mereghetti 1995).
The factor of $\sim10$ increase of the X-ray flux in the light curve measured
with
ROSAT during the monitoring of one orbital cycle (Taylor \etal/ 1996)
may be the first evidence of orbital-modulated X-ray emission.
\medskip

Despite the firm detections of \gts/ up to X-rays and of \cg/ above
100 MeV, the situation around 1 MeV is less clear.
When we compare the spectral index measured for \gts/ with
ROSAT ($\alpha\approx1.1$: Goldoni \& Mereghetti 1995) with that measured
for \cg//\groname/ with EGRET ($\alpha\approx2.21$: Fierro 1995),
it is evident that a spectral break must occur in between. This naturally also
holds if the two sources are not related.
A simple extrapolation of the measured power laws to the MeV energy range
predicts fluxes in the range of $10^{-5}-10^{-4}\photcmsmev$, which is close
to the detection threshold of COMPTEL.
A preliminary analysis of COMPTEL observations of the \cg/ field,
based on three observations made between November 1991 and August 1992, did not
reveal evidence for such a strong source but showed the presence of a weak
excess consistent with the position of \cg/ (van Dijk \etal/ 1994).

\section{Instrument and Observations}

COMPTEL (Sch\"onfelder \etal/ 1993) is a \gamms/-ray telescope on board CGRO,
sensitive in the 0.75--30 MeV range. It has a wide field of view of
$\sim1$ steradian, a source location accuracy of $\sim1^\circ$ and an energy
resolution ranging from $\sim10\%$ at 1 MeV to $\sim5\%$ at 6 MeV. Incoming
photons first Compton scatter on an electron in the upper layer of detectors
(D1) through an angle $\phigeo$ after which they are completely or partially
absorbed in the lower detector layer (D2).\par
The COMPTEL data consist of events which are defined as coincident interactions
in the two detector layers. These events are binned in a
3-dimensional data space, of which the first two dimensions denote
the coordinates $\chi$ and $\psi$ of the direction of the scattered photon.
The third dimension, the calculated scatter angle $\phibar$, follows from the
well-known Compton formula and is equal to the true scatter angle $\phigeo$
only if the electron and photon are completely absorbed in D1 and D2
respectively.\par
Given a celestial intensity distribution,
the expected distribution of events in the 3D data space is the convolution
of the intensity distribution with the instrument response function,
which is the product of a 3D point-spread function
(PSF) and a total exposure. The latter has been subdivided into a 3D
geometry function and a 2D exposure function for practical reasons.
Throughout this paper we used PSFs obtained from Monte Carlo simulations,
taking into account the full spherical response representation.

Table~\ref{tab:obslog} gives a log of the observations which were used for
the analysis presented here. Unfortunately, during
most of these observations, the angular distance between
the source and the pointing direction was not favorable.

\begin{table}
   \def\one{&} \def\two{}\def\endl{\\}
   \begin{flushleft}
   \tabcolsep=5pt
   \begin{tabular}{llcc}
   \hline
\two Obs   \one \hfil Time     \one Phase     \one $Z[{}^\circ]$  \endl
\hline
\two15      \one 1991, Nov. 28--Dec. 12   \one 0.096-0.631  \one 22.2  \endl
\two31      \one 1992, Jun. 11--Jun. 25   \one 0.499-0.023  \one 29.3  \endl
\two34      \one 1992, Jul. 16--Aug. 06   \one 0.820-0.610  \one 27.2  \endl
\two211     \one 1993, Feb. 25--Mar. 09   \one 0.272-0.725  \one 11.4  \endl
\two319.0   \one 1994, Mar. 01--Mar. 08   \one 0.197-0.464  \one 28.1  \endl
\two319.5   \one 1994, Mar. 15--Mar. 22   \one 0.727-0.990  \one 27.0  \endl
\two325     \one 1994, Apr. 26--May 10    \one 0.311-0.840  \one 15.2  \endl
\hline
\end{tabular}
\end{flushleft}
\caption{List of CGRO observation periods
during which \cg/ was within
$30^\circ$ of the pointing direction. The 4 columns give: 1) the observation
number in CGRO notation;
2) the start and end date of the observation;
3) the corresponding radio phase interval;
4) the angular distance of \cg/ to the pointing direction.
}
\label{tab:obslog}
\end{table}

\section{Analysis}

\begin{figure}[tb]
\vskip7.7truecm
\includegraphics{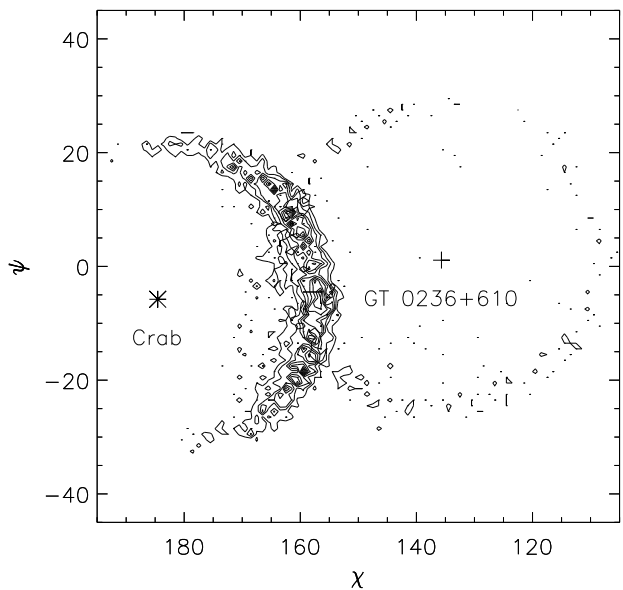}
\caption{This figure shows the $(\chi,\psi)$-distribution of events
in a data space containing simulated point sources at the position of the
Crab and at the refined location of \cg//\allowbreak\groname/,
subject to a selection on the derived scatter angle ($\phibar$) of
$[26^\circ,28^\circ]$.
In this representation, point sources show up as
ring-shaped distributions centered on the source positions with a radius
of $\phibar^\circ$.
}\label{fig:phibcut}
\end{figure}

The COMPTEL data were analysed in
the standard energy ranges 0.75--1, 1--3, 3--10 and 10--30 MeV,
as well as
in a combined 1--30 MeV energy range.
Significance and flux maps were produced using
the maximum-likelihood ratio (MLR) method.
  For a description of the application of this method to COMPTEL data,
  the reader is referred to de Boer \etal/ (1992).
  The $1\sigma$ errors quoted in this paper are statistical only, i.e. they
  do not include the estimated systematic $30\%$
  flux uncertainty.

   \subsection{Model fitting and the instrumental background model}
To account for flux contributions from nearby point sources and the underlying
Galactic diffuse emission, the MLR analysis allows for the inclusion of
{\em data-space models} in addition to an instrumental background model.
Such data-space models are created by folding the celestial
intensity distribution, a simple delta function for point sources,
through the instrument response.\par

The basic principle of the instrumental background model used here
is described by Bloemen \etal/
(1994). It is created by fitting $\phibar$-templates, obtained from the scaled
geometry function, to the event data space itself.
This background model has recently been improved upon by iteratively
correcting the $\phibar$-templates for contributions from fitted models of
detected celestial sources. The modification becomes important when
there are strong sources (e.g., the Crab or large-scale diffuse emission)
in the field of view.  In the first iteration of the model fitting, no
correction
is performed. If a positive signal is found for one or more models, a second
iteration can be performed during which the $\phibar$-templates are corrected 
for the $\phibar$-dependent model contributions. This procedure is repeated
until the scale factors of the fitted models converge.
\medskip

\begin{figure}[tb]
\vskip8.39truecm
\includegraphics{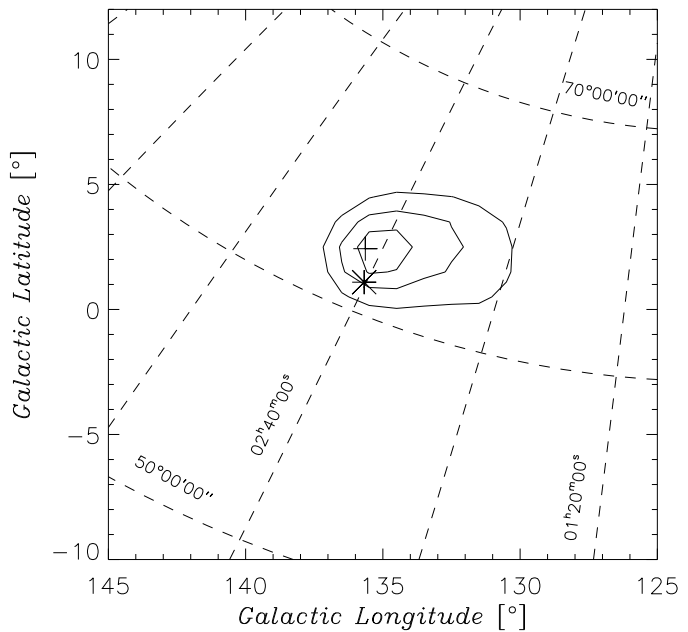}
\caption{Location contours ($1\sigma$, $2\sigma$, $3\sigma$) for the 1--30 MeV
excess for the sum of the observations listed in
Table~\protect\ref{tab:obslog}.
The positions of \gts/ and the QSO \qso/ are denoted with `\asterisk' and `+'
respectively.
}\label{fig:loccontours}
\end{figure}

The effective exposure function for the observations listed in
Table~\ref{tab:obslog} covers a
large part of the sky and extends beyond $l=200^\circ$. We therefore must take
into account the non-negligible influence of the events originating from the
Crab
($\sim2.5\times 10^{4}$ in the 1--30 MeV range compared to $\sim7\times10^{3}$
for \cg/) on the background model
by including a data-space model for the Crab.
For this set of observations, the situation is particularly confusing
because the signals of the Crab and of a source at the position of
\cg//\groname/ are at a maximum in the same region of the data space
(the region where the geometry function peaks, see Fig.~\ref{fig:phibcut}).
When models for both the Crab and for \cg//\groname/ are included
simultaneously,
this sometimes causes the Crab flux to be underestimated and the
\cg//\groname/ flux
to be overestimated in the first iteration of the background modelling.
In general, $\sim4$ iterations are needed to correct for this effect.

\section{Results}

\subsection{Spatial analysis}

\label{sec:spatialana}

When all the observations in Table~\ref{tab:obslog} are summed, an excess
consistent with the position of \cg/ is found at energies above 1 MeV. Below
1~MeV an upper limit is obtained.  Fig.~\ref{fig:loccontours} shows
the location contours for the 1--30 MeV energy range; the maximum of the
excess has a MLR value of 50.9. This corresponds to a point-source detection
significance of $6.6\sigma$ for 3 d.o.f. (neglecting the possible presence of
underlying Galactic diffuse emission), while at the position of \cg/ it is
$6.7\sigma$ (1 d.o.f.). In the standard energy ranges 1--3 MeV, 3--10 MeV
and 10--30 MeV, the peak MLR values are 24.4, 18.8 and 10.3 respectively.
In the individual observation periods the signal is either weak or not detected
at all.\par

\begin{figure*}[tb]
\vskip10.5truecm
\includegraphics{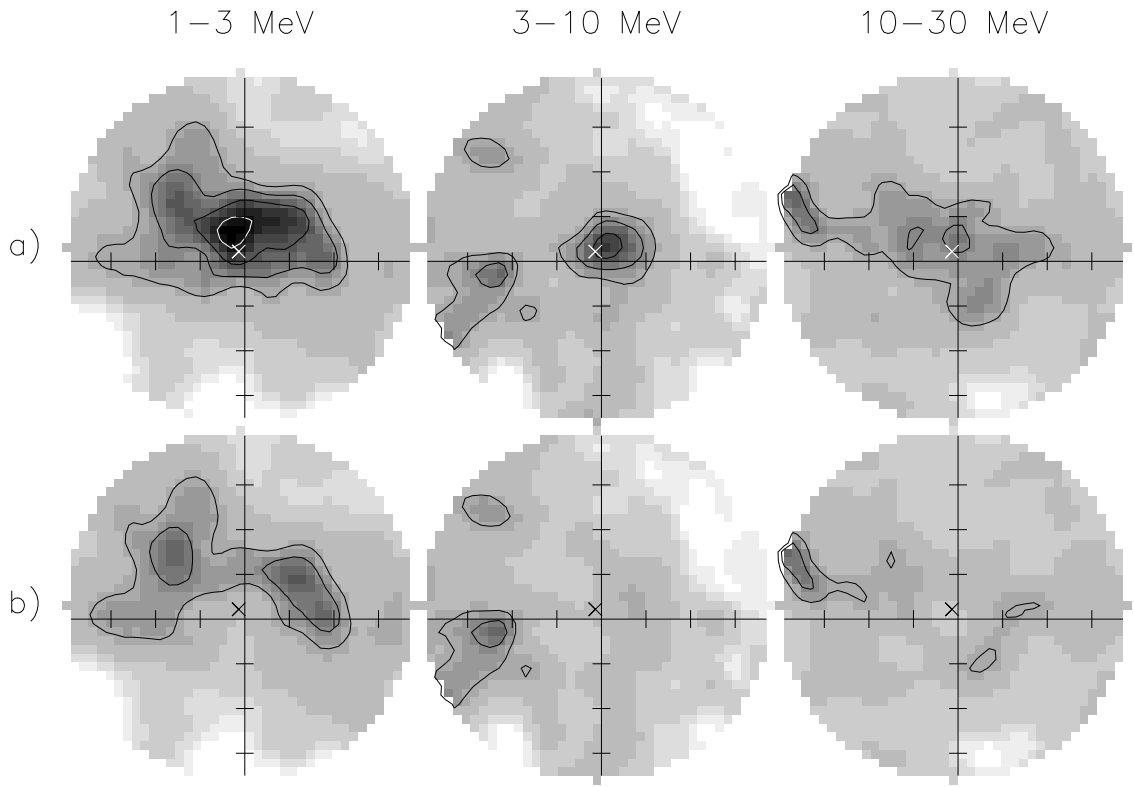}
\caption{MLR maps for the region around \cg/, centered on
$(l,b)=(\decree{135}{.5},\decree{1}{.5})$
with a radius of $20^\circ$, longitude $l$ increasing towards the left.
The horizontal and vertical lines are drawn at $b=0^\circ$ and $l=135^\circ$
respectively and the position of \cg//\groname/ is denoted by the `$\times$'
symbols.
Contour levels are spaced as 4, 9, 16 and 25.
{\em a)} including a model for the Crab;
{\em b)} including models for both the Crab and for \cg//\groname/.
}\label{fig:grmlrmaps}
\end{figure*}
\medskip

In Fig.~\ref{fig:grmlrmaps}a, the significance maps for the standard
energy ranges above 1~MeV are shown. In each of these maps, an excess
consistent with the position of \cg/ can be observed. The significance
in the highest
energy range is marginal ($2.6\sigma$), while below 3 MeV the excess is broader
than expected for a single point source.
When a data-space model for a source at the position
of \cg//\groname/ is included, we obtain the residual maps as shown in
Fig.~\ref{fig:grmlrmaps}b.
None of the remaining excesses by itself is
formally significant at the $\ge3\sigma$ level for a free search, but the
number of $2\sigma-3\sigma$ excesses yield it
unlikely that all are just statistical fluctuations.
One of the possible celestial components, that of the Galactic diffuse
emission, is discussed in Section~\ref{sec:galdif}. A possible contribution
from point sources, however, cannot be excluded. Most noticeable is the
weak emission around $(l,b)=(\decree{134}{.6},\decree{+7}{.0})\equiv P_1$
in the 1--3 MeV map in Fig.~\ref{fig:grmlrmaps},
which is completely due to a strong ($3.9\sigma$) excess observed during
Obs.~325. It is not seen in any other energy range nor in any
other COMPTEL observation (see also Bhattacharya \& Owens 1994).
After a correction for a minimum number of
trials (4 energy intervals, 7 observations), its significance in Obs.~325
is only slightly larger than $3\sigma$, which does not confidently exclude
a statistical background fluctuation as origin.
However, located only $6^\circ$ away from \cg//\groname/ (but not consistent
with the position of
the latter at the $>3\sigma$ level), the apparent source $P_1$ has
a non-negligible influence on the flux derived for the COMPTEL source
in the mosaic of the observations and thus has to be accounted for.

\subsection{Spectral analysis}

\label{sec:specana}

The analysis in the previous section was performed with
PSFs for an assumed $E^{-\alpha}$ power-law photon spectrum with $\alpha=2$.
Since the fluxes and
likelihoods depend only slightly on the assumed input spectrum, this was not
a bad approximation. In order to find the power-law index that fits
the data best, the MLR analysis has been repeated using PSFs with various values
for $\alpha$ until the ratios of the fluxes measured in different energy
ranges are consistent with the assumed input spectrum. \par
Such flux ratios were calculated using PSFs with indices ranging from
1.4 to 2.8. In the model fitting we included a data-space model for the
Crab in each energy range, while for the 1--3 MeV range we also included a
model for the nearby excess P1 to correct for its influence on the flux
derived for the COMPTEL source.
In Table~\ref{tab:ratios} we show the results for the extreme values of
index $\alpha$ as well as for $\alpha=2$.  Also listed are the {\sl expected}
ratios for power-law spectra with these indices.

\begin{table}
   \def\one{&} \def\two{}\def\endl{\\}
   \begin{flushleft}
   \tabcolsep=3pt
   \begin{tabular}{c|c|c|c}
   \hline
\two Index \one $R_A$ (1-3/3-10) \one $R_B$ (3-10/10-30) \one $R_C$ (1-3/10-30)
   \endl
   \hline
\two 1.4     \one $2.3\pm0.8$  \one $4.1\pm1.6$  \one $9.4\pm3.8$    \endl
\two         \one 1.44         \one 1.74         \one 2.51           \endl
   \hline
\two 2.0     \one $2.2\pm0.8$  \one $4.1\pm1.6$  \one $9.1\pm3.8$    \endl
\two         \one 2.86         \one 3.50         \one 10.00          \endl
   \hline
\two 2.8     \one $2.1\pm0.8$  \one $4.2\pm1.6$  \one $8.7\pm3.8$    \endl
\two         \one 7.03         \one 8.98         \one 63.10          \endl
   \hline
   \end{tabular}
   \end{flushleft}
\caption{Measured (first line of each entry) and expected (second line of
each entry) hardness ratios for power-law
input spectra with different spectral indices. These hardness ratios are
defined as the ratio of the fluxes in the specified energy ranges
(units [$\hbox{photons cm}^{-2}\hbox{ s}^{-1}$]).
}\label{tab:ratios}
\end{table}

\begin{table}
   \def\one{&} \def\two{}\def\endl{\\} \def\form{&\multispan2}
   \begin{flushleft}
   \tabcolsep=5pt
   \begin{tabular}{l|c|c}
   \hline
   \noalign{\vskip0.04truecm}
\two $E$ [MeV] \form \hfil Flux
           $[\hbox{photons cm}^{-2}\hbox{ s}^{-1}]$\hfil  \endl
   \hline
\two       \one no diff. models \one incl. diff. models  \endl
   \hline
\two 0.75--1  \one $<4.8\times10^{-5}$    \one $<4.8\times10^{-5}$ \endl
\two 1--3   \one $(11.2\pm3.2)\times10^{-5}$ \one $(6.0\pm3.0)\times10^{-5}$
      \endl
\two 3--10  \one $(5.0\pm1.2)\times10^{-5}$  \one $(2.2\pm1.2)\times10^{-5}$
      \endl
\two 10--30 \one $(1.2\pm0.4)\times10^{-5}$  \one $(0.3\pm0.4)\times10^{-5}$
      \endl
   \hline
   \end{tabular}
   \end{flushleft}
\caption{Time-averaged fluxes with $1\sigma$ errors and the $2\sigma$ upper
limits for the COMPTEL source consistent with the position of \cg/, for the
sum of observations given in Table~\protect\ref{tab:obslog}.
These results were derived including a model for
the Crab in all energy ranges. In the 1--3 MeV range a
point-source model at $(l,b)=(\decree{134}{.6},\decree{7}{.0})$ was
included as well (see text). The second column shows the results obtained
disregarding the possible presence of underlying diffuse emission, the
third column shows the results when models for the Galactic diffuse
emission are incorporated (see Section~\protect\ref{sec:galdif}).
}\label{tab:fluxes}
\end{table}

Although the measured ratios are not very sensitive to the assumed
input spectrum, the expected ratios differ considerably. The indices
that fit the COMPTEL data best are $\alpha_0=1.8^{+0.3}_{-0.4}$
and $\alpha_0=2.15^{+0.3}_{-0.4}$ for the $R_A$ and $R_B$ hardness
ratios respectively. They
are consistent with a single power law $F=I(E/1\hbox{ MeV})^{-\alpha_0}$ across
the entire COMPTEL energy range with $\alpha_0=1.95^{+0.2}_{-0.3}$ and
amplitude $I=1.75\times10^{-4}\photcmsmev$.
Adopting this spectral index, the time-averaged COMPTEL fluxes
and upper limits are given in the second column of Table~\ref{tab:fluxes}.
The upper limit
in the 0.75--1 MeV range is inconsistent with the power law at the
$2.4\sigma$ level, which suggests that there is a spectral
break in the COMPTEL energy domain. Note that at 1 MeV, the COMPTEL flux levels
are a factor of $10-100$ below those reported in the past for the emission
from this celestial region
(Coe \etal/ 1978; Perotti \etal/ 1980).

\subsection{Temporal analysis}

\label{sec:tempana}

We performed a search for flux variations correlated with the
orbital period of \gts/. For this the data for each energy range were binned
and analysed in 5 phase bins of width 0.2, assuming an orbital period of
$26.496\pm0.008$ days with phase 0 at JD 2,443,366.775 (Taylor \& Gregory
1984). In the phase bins that contained a contribution
from Obs. 325, we again included a point-source model for the $P_1$ excess
observed in that observation in the 1--3 MeV range (see
Section~\ref{sec:spatialana}).
The radio lightcurves differ significantly from
cycle to cycle and the phase of the peak outburst is seen to vary by more than
$\sim0.5$. Therefore, this approach may fail to detect correlations of the weak
\gamms/-ray flux with the orbital period if the \gamms/-ray flux variations
occur, like the radio-flux variations, at irregular phases.
Note that the error on the orbital period corresponds to a total relative phase
shift during the observations given in Table~\ref{tab:obslog} of only 1\%, but
that the absolute phase may
be wrong by $\sim6\%$, corresponding to $\sim1.5\hbox{ days}$.
\begin{figure}[tb]
\vskip5.5truecm
\includegraphics{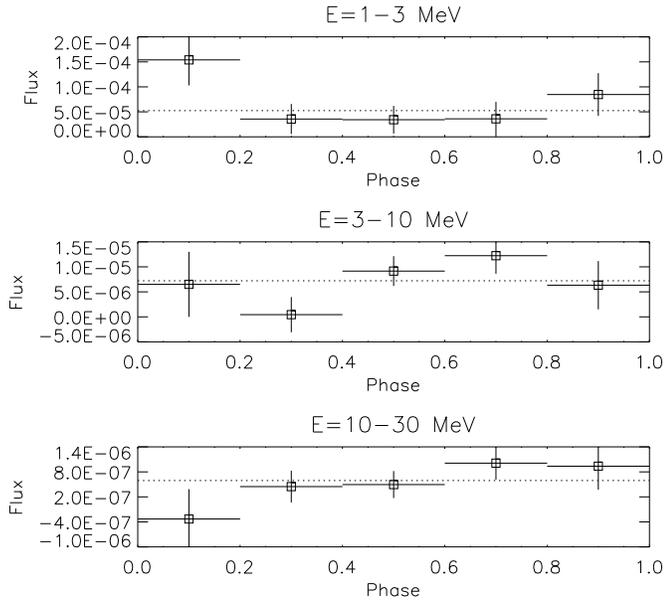}
\vskip2.2truecm
\caption{Flux $[\photcmsmev]$ as a function of orbital phase for three energy
ranges. The dashed line shows the result from a $\chi^2$-fit with a constant
distribution.
}\label{fig:temporal}
\end{figure}

The results of this analysis are shown in Fig.~\ref{fig:temporal}. In each
energy range, a $\chi^2$-fit with
a constant distribution has been used to determine if there is evidence for
orbital-phase related flux variations. The $\chi^2$ values thus obtained
are 5.49, 6.10 and 3.38 for 4 d.o.f., indicating that the COMPTEL data do
not show evidence for flux variations correlated with the orbital phase.
Due to the large statistical flux errors, however,  and the above-mentioned
possibility of peak fluxes occurring at irregular phases, \gamms/-ray
variability similar to that observed in X-rays with ROSAT (Taylor \etal/ 1996)
cannot be excluded.

\subsection{The Galactic diffuse emission}

\label{sec:galdif}

So far in the analysis we have not taken into account the possible
contribution of diffuse emission from the Galactic disk. If this component,
which is dominant at energies above 100 MeV, is indeed not negligible
at COMPTEL energies in this region of the sky, then the flux of the COMPTEL
source could be significantly lower than derived above.
Furthermore,
Monte Carlo simulations have shown that weak diffuse emission from a
Galactic-disk distribution may result in several weak point-source-like
excesses at low latitudes, similar to those observed in
Fig.~\ref{fig:grmlrmaps}.

\subsubsection{The diffuse models}

The classical interpretation of the Galactic diffuse emission above 100
MeV is that it is dominated by $\pi^0$-decay,
resulting from nuclear interactions between cosmic-ray (CR)
particles and
nuclei of the interstellar gas (for an overview see Bloemen 1989).
Below 100 MeV, the main contribution is
thought to come from bremsstrahlung losses of low-energy CR electrons
interacting with the interstellar gas. In both cases, the celestial
\gamms/-ray distribution is expected to correlate approximately with that of
the interstellar gas, as mapped by HI and CO surveys (the latter component
assumed to be a tracer of $\molh$). Studies performed
with SAS-2 and COS-B data ($\ga100\hbox{ MeV}$) have shown that such a
correlation indeed exists (e.g., Fichtel \& Kniffen 1984; Bloemen 1989).
It should be noted that the assumptions about the CR density distribution,
e.g., whether or not it is coupled to matter, differ among the various studies.
Another component ($\equiv\hbox{IC}$) that is thought to contribute to the
Galactic diffuse
emission both below and above 100 MeV is inverse Compton radiation from
high-energy CR electrons scattering on the interstellar radiation field.
The relative contribution of this component near the Galactic plane probably
does not exceed 5\%--10\% at energies near 100 MeV (see Bloemen, 1989, for an
overview).

Recent analyses revealed that the COMPTEL data show evidence for a
broader latitude distribution than those of the HI and CO models.
It remains uncertain, however, whether this should indeed be attributed
to IC emission. Furthermore, an emission component that correlates with
the CO distribution appears to be lacking in the COMPTEL energy
regime (Strong \etal/ 1996; Bloemen \etal/ 1996), except perhaps for the 10-30
MeV energy range (van Dijk 1996).
Also, it should be noted that part of the inferred diffuse intensities may
be due to unresolved, weak point sources.
Clearly, the problem of the Galactic diffuse emission at
COMPTEL energies requires further study. Here we will use the current
understanding to discuss the possible implications for the quantities derived
in Sections~\ref{sec:specana} and \ref{sec:tempana}.

\subsubsection{Implications for derived quantities}

\label{sec:implicat}

In order to assess the possible influence of the Galactic diffuse emission
underneath the COMPTEL source on the derived source flux, we repeated the MLR
analysis presented above including data-space models for the interstellar gas,
using the same HI and CO surveys that were used in the COS-B studies (Bloemen
1989) and adapting the model for the IC distribution from Strong \& Youssefi
(1995). Firstly, we fitted only the HI and IC models, fixing the HI
emissivities and the scaling factors for the IC models at the values found by
Strong \etal/ (1996).
As a result, the fluxes in the 1--3 MeV and 3--10 MeV ranges are reduced by a
factor of $\sim2$ (third column in Table~\ref{tab:fluxes}). Above 10 MeV,
the flux drops below $1\sigma$ significance. The photon
spectral index derived from these reduced 1--3 MeV and 3--10 MeV fluxes
is consistent with that derived in Section~\ref{sec:specana}, although not
very constraining.
Since the errors on the fluxes have remained roughly constant,
the temporal analysis from Section~\ref{sec:tempana} is expected to yield even
lower $\chi^2$-values and was therefore not repeated.

Secondly, we investigated a worst-case scenario in which the CO model is added
to the HI and IC models with a fixed value for the CO-to-$\molh$ conversion
factor $X_\gamma=2.0$. Although
correlations of the $\gamma$-ray emission with the CO distribution have not
been found for COMPTEL data, fixing $X_\gamma$ at the value found above 100
MeV gives an estimate of the error made in case the CO component is
present, but for some reason is not recognized. The inclusion of the CO model
is found to result in a small decrease of the 1--3 MeV and 3--10 MeV fluxes
inferred for the COMPTEL source of less than $\sim15\%$.
\medskip

The effect of including the models for the Galactic diffuse emission on the
parameters derived for the COMPTEL source can only be
considered to be crudely estimated.
It is therefore likely that the fluxes given in the second column of
Table~\ref{tab:fluxes} are overestimated, but due to the uncertainties in
modelling the Galactic diffuse emission, the
fluxes given in the third column of Table~\ref{tab:fluxes} should be
interpreted with caution as well.

\section{Discussion}

\label{sec:disc}

\subsection{Identification of the COMPTEL source}

\begin{figure*}[tb]
\vskip12.0truecm
\includegraphics{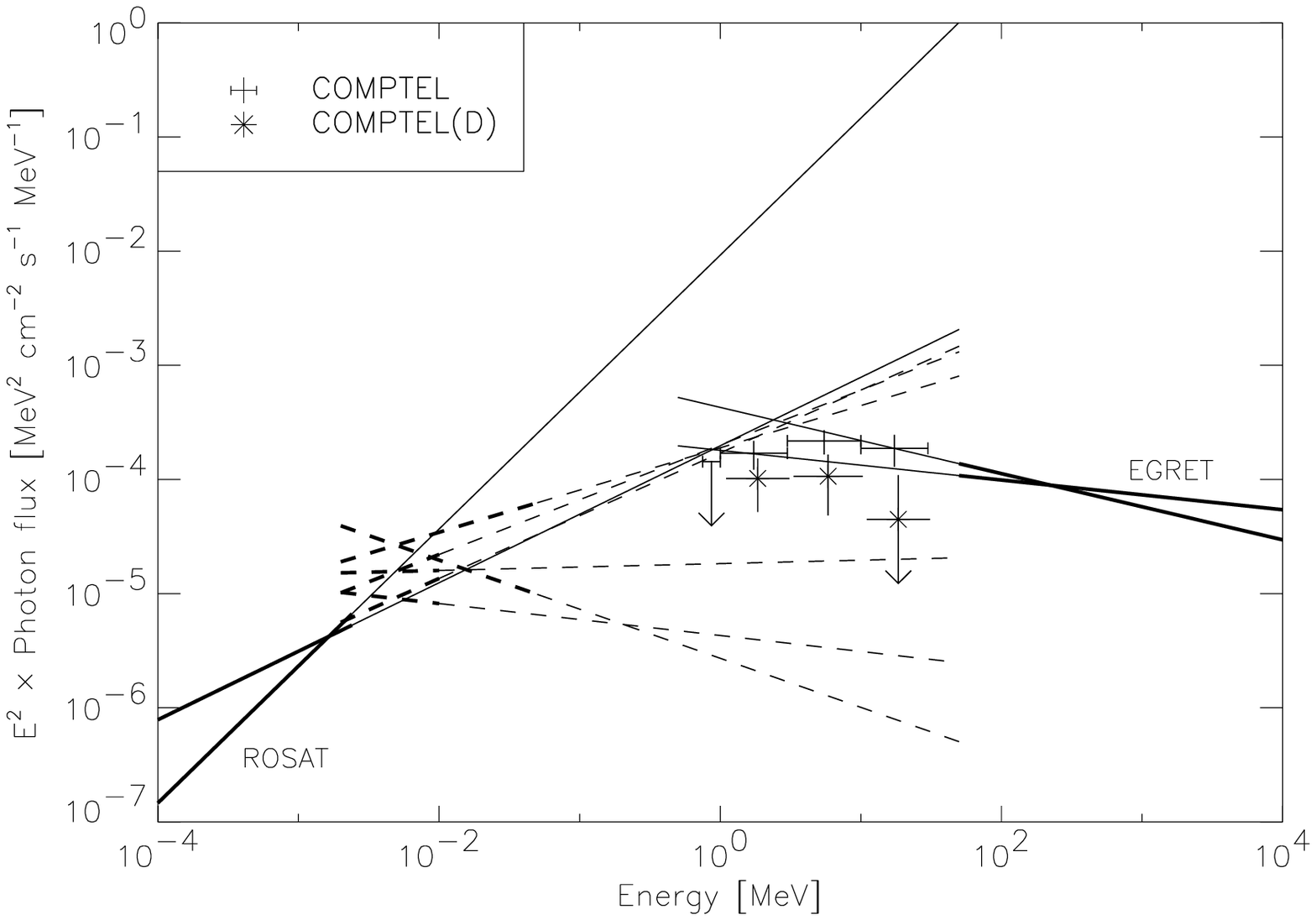}
\caption{The phase-averaged fluxes $(\times E^2)$ for the COMPTEL excess `as
is' (points labeled `COMPTEL'; second column in
Table~\protect\ref{tab:fluxes}), and those corrected for the estimated
contribution from the Galactic diffuse emission (points labeled `COMPTEL(D)';
third column in Table~\protect\ref{tab:fluxes}),
compared with unabsorbed X-ray and \gamms/-ray data for \gts/
(lines labeled `ROSAT'; Goldoni \& Mereghetti 1995),
\cg//\groname/ (lines labeled `EGRET'; Fierro 1995) and \qso/ (dashed lines).
All data shown, except for the COMPTEL data, are extrapolated power laws,
from the energy range in which the
measurements were made (thick lines), into the MeV range (thin lines). In
each case, two power laws are shown,
reflecting the uncertainties in the photon spectral indices. The X-ray
data for \qso/ (dashed lines) were  measured with
Einstein (Turner \etal/ 1991),
OSO 8 (Worrall \etal/ 1980) and
EXOSAT (Turner \& Pounds 1989).
}\label{fig:compspec}
  \end{figure*}

The celestial distributions of neither the HI nor the IC models
can mimic a
point-source-like feature at the position of \cg/. So, despite the
uncertainties in the absolute
flux levels of the COMPTEL source due to the unknown contribution of the
Galactic diffuse emission, the mere presence of a source
consistent with the position of \cg/ is evident.
If we extrapolate the time-averaged high-energy \gamms/-ray spectrum
measured for \cg//\groname/ (Fierro 1995; lines labeled `EGRET'
in Fig.~\ref{fig:compspec}), we find that the COMPTEL fluxes
prior to the correction for the Galactic diffuse emission
(points labeled `COMPTEL' in Fig.~\ref{fig:compspec}) are consistent
with the extrapolations. When the models for the
Galactic diffuse emission are included, the fluxes are still consistent, but
appear to be somewhat low (points labeled `COMPTEL(D)'). Although this
may indicate that the contribution of the diffuse emission is overestimated,
a spectral break is expected to occur somewhere in this energy range.

The extrapolations of the X-ray spectra measured for \gts/ (lines labeled
`ROSAT'
in Fig.~\ref{fig:compspec}) and for the quasar \qso/ (the dashed lines)
do not reject any of the two being the possible counterpart of the COMPTEL
source.  There are several
arguments, however, which favour \gts/ being the counterpart.
Firstly, \qso/ has only been detected up to $\sim20\hbox{ keV}$
(Turner \& Pounds, 1989) and is not seen with EGRET $(>30\hbox{ MeV})$. On the
other hand, both ROSAT and EGRET have reported detections coincident with \gts/,
with location uncertanties that leave little room for other counterparts
(Goldoni \& Mereghetti 1995; von Montigny \etal/ 1993; Thompson \etal/ 1995).
Secondly, the only active galactic nuclei (AGNs) that have been detected with
COMPTEL (and EGRET) up to now are those which are radio loud
and have flat radio spectra. This is consistent with the findings from an
analysis of EXOSAT ME data by Lawson \etal/ (1992), who found a steeper
average 2--10 keV spectrum for radio-quiet QSOs (photon index
$\bar\alpha=1.9\pm0.11$) than for radio-loud QSOs ($\bar\alpha=1.66\pm0.07$).
Although \qso/ has been shown to contain
a compact radio core with an inverted radio spectrum, the radio flux
of $\sim 0.20\hbox{ Jy}$ at 5 GHz is rather low (Apparao \etal/ 1978) and
has only been detected at radio wavelengths because it is relatively nearby
($z=0.0438$).
Unless \qso/ behaves very differently from other radio-quiet QSOs at MeV
energies, it is not likely that its spectrum extends into the COMPTEL 
energy range.
\medskip

Apart from \gts/ and \qso/, there are no other sources in the SIMBAD data base
that can be regarded potential counterparts, in the sense that there are no
strong non-thermal X-ray or high-energy \gamms/-ray emitters nearby.
A possible contribution from the large molecular complex W3 at
$(l,b)\approx(134^\circ,1^\circ)$, as part of the Galactic diffuse emission,
was estimated to have a small impact (Section~\ref{sec:implicat}).

\subsection{Comparison with theory}

The previous sections have shown that the identification of the COMPTEL source
with \cg/ is likely, but that the absolute flux levels are
uncertain by a factor of two. Here we will address
the possible implications for the main theoretical models for \cg//\gts/,
assuming for now that the radio source and the \gamms/-ray source are indeed
one and the same.
\medskip

Based on the presence of a $\sim1.4\solmas$ compact object (Hutchings \&
Crampton 1981), two basic models
exist to explain the variable radio emission and the possible high-energy
emission of the Be/X-ray binary \gts/.
In the {\em supercritical accretion\/} (SA) model
(Taylor \& Gregory 1982, 1984; Taylor \etal/ 1992), a neutron star is in
an eccentric orbit around the Be star, accreting from the slow high-density
equatorial wind of the latter. For suitable chosen values of the wind
parameters and the eccentricity, the radio outbursts arise from
the energetic particles produced during the supercritical accretion that
occurs near periastron passage. The four-year modulation and the varying
phases of the peak radio fluxes may be explained by a varying
speed of the equatorial wind of the Be star (Mart\'\i\ \& Paredes 1995).
However, it is not clear why the X-ray luminosity during outbursts is several
orders of magnitudes below that for other Be/X-ray binaries
(Taylor \etal/ 1996).

Maraschi \& Treves (1981), on the other hand, proposed the so called
{\em young-pulsar\/}
model, in which the outbursts and high-energy emission originate in the shock
front that results from the collision of the stellar-disk wind and the
relativistic pulsar wind. Such a model has also been used for the highly
non-thermal
1--200 keV emission detected from the Be star/pulsar system PSR~1259--63
containing a 47 ms radio pulsar (Tavani, Arons \& Kaspi 1994; Tavani \etal/
1996). In the case of the PSR~1259--63 system, accretion onto the
surface of the neutron star is inhibited by the strong radiation pressure
from the pulsar's electromagnetic and relativistic particle wind. The
high-energy emission is produced by synchrotron and inverse-Compton
emission of shock accelerated particles. Shock-powered high-energy
emission is of general importance in pulsar binaries and
the \cg/ system may belong to the category of hidden pulsars
(Tavani 1995).
\medskip

Both types of models involve a neutron star in an eccentric orbit around
the Be star and predict, to some degree, time variability of the \gamms/-ray
emission correlated with the orbital period. The COMPTEL data on the other
hand are consistent with a constant
flux, but the large statistical errors on the COMPTEL fluxes
(Table~\ref{tab:fluxes}) do not allow for tight constraints on the variability.
At \gamms/-ray energies above 100 MeV, the chance that the flux from
\cg//\groname/ is constant between observations is only 0.1\% (Fierro 1995),
while an increase by a factor of $\sim2$ was observed within a
single observation (Kniffen \etal/, in preparation).
In the young-pulsar model of emission,
orbital modulation of the high energy emission depends on the poorly
understood pitch angle evolution of radiating shock-accelerated particles
of the pulsar wind. Upstream of the shock, the particles flow radially away
from the pulsar
and therefore initially have a nonisotropic velocity
distribution. For large values of the ratio of the electromagnetic and plasma
flow energies upstream, the shocked particles acquire preferential directions
of motion (Gallant \etal/ 1992) and a dependence of the \gamms/-ray emission
on orbital phase might be observed. If, on the other hand, this ratio is low,
the onset of the Weibel instability may isotropize the downstream flow and
therefore the \gamms/-ray emission (Weibel 1959; Hoshino \etal/ 1992).
Note that in both of these cases,
the shock-powered emission may be more dependent on
the varying conditions of the pulsar cavity being compressed by the
gaseous outflow from the companion star (Tavani, Arons \& Kaspi 1994).
\medskip

The COMPTEL measurements presented here imply a luminosity in the
1--30 MeV energy range of
$5.0\times10^{35}(D/2\hbox{ kpc})^2\ergss$
($2.0\times10^{35}\ergss$
if the models for the diffuse emission are included). Together with the ROSAT
and EGRET measurements, this implies a total luminosity above 0.1
keV of $L\approx1.5\times10^{36}\ergss$. For comparison, the inferred
total energy in the relativistic electrons responsible for the radio outbursts,
which typically last several days, is only $\sim2\times10^{39}\hbox{ ergs}$
(Taylor \& Gregory 1984).
A high-energy luminosity near $10^{36}\ergss$
may be powered by shock
emission of a relativistic pulsar wind, as observed in the case of
the plerionic Crab Nebula and for the binary Be star/pulsar system
PSR~B1259--63 (Tavani \etal/ 1996).
Here, the deduced spindown luminosity of the underlying
pulsar would be of the order of $10^{37}\ergss$ for a conversion
efficiency into the \gamms/-ray band of about 10\% (similar to the
conversion efficiencies observed for the Crab Nebula and for
PSR~B1259--63). The possible spectral break observed for \cg/ in the MeV
energy range may, when more accurately determined,
be used to constrain parameters such
as the kinetic energy of the particles and the electromagnetic energy
flux upstream.

\section{Conclusions}

We have detected a COMPTEL source consistent with the
position of \cg//\groname/
in the energy range 1--30 MeV. The flux at 1 MeV is roughly two
orders of magnitude below the values reported in the past for a source in this
celestial region. The COMPTEL spectral shape
is consistent with the spectral break expected from extrapolations of
the ROSAT and EGRET observations of \gts/ and \cg//\groname/ respectively.
\gamms/-Ray observations with higher spatial resolution, or showing correlated
variability, are necessary to  prove unambigously the association
between the radio source \gts/ and the \gamms/-ray source
\cg/.

\acknowledgements{We thank J.~van Paradijs for his helpful comments on an
earlier version of the manuscript.
We acknowledge the constructive comments from the referee G.F.~Bignami,
which have improved the readability of the
paper. R.~van Dijk acknowledges the support by the Netherlands
Foundation for Research in Astronomy (NFRA), with financial aid from the
Netherlands Organisation for Scientific Research (NWO).
The COMPTEL project is supported by the German government
through DARA grant
50 QW 90968, by NASA under contract NAS5-26645 and by the Netherlands
Organisation for Scientific Research (NWO).
M.~Tavani acknowledges the support by the NASA grant no.~NAG 5-2729.
This research has made use of data obtained through the High Energy
Astrophysics Science Archive Research Center Online Service, provided by the
NASA-Goddard Space Flight Center.
This research has also made use of the Simbad database, operated at CDS,
Strasbourg, France.}


\begin{thebibliography}{}
\bibitem{} Apparao K.M.V.~\etal/, 1978, Nat 273, 450
\bibitem{} Bhattacharya D., Owens A., 1994, ApJ 430, 371
\bibitem{} Bignami G.F.~\etal/, 1981, ApJ 247, L85
\bibitem{} Bloemen H., 1989, ARA\&A 27, 469
\bibitem{} Bloemen H.~\etal/, 1994, ApJS 92, 419
\bibitem{} Bloemen H.~\etal/, 1996, in preparation
\bibitem{} D'Amico N.~\etal/, 1987, A\&A 180, 114
\bibitem{} de Boer H.~\etal/, 1992, in: Data Analysis in Astronomy IV, eds.~V.
           Di Ges\`u \etal/ (New York: Plenum Press), Vol 59, p241
\bibitem{} Coe M.J., Quenby J.J., Engel A.R., 1978, Nature 274, 343
\bibitem{} Fichtel C.E., Kniffen D.A., 1984, A\&A 134, 13
\bibitem{} Fierro J.M., 1995, Ph.D.~Thesis, Stanford University
\bibitem{} Gallant Y.A.~\etal/, 1992, ApJ 391, 73
\bibitem{} Goldoni P., Mereghetti S., 1995, A\&A 299, 751
\bibitem{} Gregory P.C., Taylor A.R., 1978, Nat 272, 704
\bibitem{} Gregory P.C.~\etal/, 1989, ApJ 339, 1054
\bibitem{} Hermsen W.~\etal/, 1977, Nat 269, 494
\bibitem{} Hoshino M.~\etal/, 1992, ApJ 390, 454
\bibitem{} Howarth I.D., 1983, MNRAS 203, 801
\bibitem{} Hunt L.K., Massi M., Zhekov S.A., 1994, A\&A 290, 428
\bibitem{} Hutchings J.B., Crampton D., 1981, PASP 93, 486
\bibitem{} Lawson A.J.~\etal/, 1992, MNRAS 259, 743
\bibitem{} Maraschi L., Treves A., 1981, MNRAS 194, 1P
\bibitem{} Maraschi L., Tanzi E.G., Treves A., 1981, ApJ 248, 1010
\bibitem{} Mart\'\i\ J.~\& Paredes J.M., 1995, A\&A 298, 151
\bibitem{} Massi M.~\etal/, 1993, A\&A 269, 249
\bibitem{} Mendelson H., Mazeh T., 1989, MNRAS 239, 733
\bibitem{} Paredes J.M., Figueras F., 1986, A\&A 154, L30
\bibitem{} Paredes J.M., Estalella R., Rius A., 1990, A\&A 232, 377
\bibitem{} Paredes J.M.~\etal/, 1994, A\&A 288, 519
\bibitem{} Perotti F.~\etal/, 1980, ApJ 239, L49
\bibitem{} Sch\"onfelder \etal/, 1993, ApJS 86, 657
\bibitem{} Strong A.W., Youssefi G., 1995, Proc.~24th ICRC, 3, 48
\bibitem{} Strong A.W.~\etal/, 1996, in proceedings of the 3rd Compton
           Symposium (Munich), in press
\bibitem{} Swanenburg B.N.~\etal/, 1981, ApJ 243, L69
\bibitem{} Tavani M., 1995, in {\it The Gamma-Ray Sky with GRO and SIGMA},
eds.~M.~Signore, P.~Salati \& G.~Vedrenne (Dordrecht: Kluwer), p.~181
\bibitem{} Tavani M., Arons J., Kaspi V.M., 1994, ApJ 433, L37
\bibitem{} Tavani M.~\etal/, 1996, A\&AS, in press
\bibitem{} Taylor A.R., Gregory P.C., 1982, ApJ 255, 210
\bibitem{} Taylor A.R., Gregory P.C., 1984, ApJ 283, 273
\bibitem{} Taylor A.R.~\etal/, 1992, ApJ 395, 268
\bibitem{} Taylor A.R.~\etal/, 1996, A\&A 305, 817
\bibitem{} Thompson D.J.~\etal/, 1995, ApJS 101, 259
\bibitem{} Turner T.J., Pounds K.A., 1989, MNRAS 240, 833
\bibitem{} Turner T.J.~\etal/, 1991, ApJ 381, 85
\bibitem{} van Dijk R.~\etal/, 1994, AIP Conf.~Proc.~304, 324
\bibitem{} van Dijk R., 1996, Ph.D.~Thesis, University of Amsterdam
\bibitem{} von Montigny C.~\etal/, 1993, IAU Circ.~5708
\bibitem{} Weibel E.S., 1959, Phys.~Rev.~Lett.~2, 83
\bibitem{} Worrall D.M.~\etal/, 1980, ApJ 240, 421

\end{thebibliography}
\end{document}